\documentclass[12pt]{article} 
\usepackage{cite}
\usepackage{wrapfig}
\usepackage{amssymb}
\usepackage{amsfonts}
\usepackage{authblk}
\usepackage{amsmath}
\usepackage{lineno}
\usepackage{color}
\usepackage{xcolor}
\usepackage{multirow}
\usepackage{longtable}
\usepackage{lscape}
\usepackage[T2A]{fontenc}
\usepackage[utf8x]{inputenc}
\usepackage{hyperref}
\usepackage{bigstrut}

\usepackage{ifpdf}
\ifpdf
  \usepackage[pdftex]{graphicx}
  \usepackage{epstopdf}
\else
  \usepackage{graphicx}
\fi
\DeclareGraphicsExtensions{.png,.pdf,.jpg,.mps,.eps}
\graphicspath{{./figs/}}
\def\bq  {\begin{equation}}
\def\eq  {\end{equation}}
\def\bqa {\begin{eqnarray}}
\def\eqa {\end{eqnarray}}
\def\nll {\nonumber\\}

\def\mz  {M_{\sss{Z}}}

\newcommand{\sss}[1]{\scriptscriptstyle{#1}}

\newcommand{\ds }{\displaystyle}
\def\alr{A_{\rm LR}}
\def\afb{A_{\rm FB}}
\def\alrfb{A_{\rm LRFB}}
\def\ptau{P_{\tau}}
\def\ael{A_{e}}
\def\ata{A_{\tau}}

\begin{document}

\title{Asymmetries in Processes of Electron--Positron Annihilation}

\author[1,2]{Andrej Arbuzov}
\author[1]{Serge Bondarenko}
\author[3]{Lidia Kalinovskaya}

\affil[1]{\small Bogoliubov Laboratory of Theoretical Physics, JINR, 
                 141980 Dubna, Moscow region, Russia}
\affil[2]{\small Dubna State University,  
                 Universitetskaya str. 19,  141982 Dubna, Russia}
\affil[3]{\small Dzhelepov Laboratory of Nuclear Problems, JINR,  
                 141980 Dubna, Moscow region, Russia}

\date{}

\maketitle

\abstract{
Processes of electron--positron annihilation into a pair of fermions were considered.
Forward--backward and left--right asymmetries were studied, taking into account
polarization of  initial and final particles. Complete 1-loop electroweak radiative 
corrections were included. A wide energy range including the $Z$ boson peak
and higher energies relevant for future $e^+e^-$ colliders was covered. Sensitivity of
observable asymmetries to the electroweak mixing angle and fermion weak coupling 
was discussed. \\[.3cm]
}

\section{Introduction}

Symmetries play a key role in the construction of physical theories. In fact, they
allow us to describe a huge variety of observables by means of compact formulae.
We believe that the success of theoretical models based on symmetry principles
is due to the presence of the corresponding properties in Nature. The Standard Model (SM)
is the most successful physical theory ever. Its~predictions are in excellent agreement
with practically all experimental results in particle physics. The~renormalizability of
the model allows us to preserve unitarity and provide finite verifiable results.
Both~phenomenological achievements and nice theoretical features of the SM are mainly
due to the extended usage of symmetries in its construction. The model is based on
several symmetries of different type, including the Lorentz (Poincar\'e) symmetry,
the gauge $SU(3)_C \times SU(2)_L \times U(1)_Y$ symmetries, the CPT symmetry,
the spontaneously broken global $SU(2)_L \times SU(2)_R$ symmetry in the Higgs sector, etc.
Some symmetries of the model are exact (or seem to be exact within the present precision)
while others are spontaneously or explicitly broken. In particular, the nature of
the symmetry among the three generations of fermions is one of the most serious puzzles
in the SM and verification of the lepton universality hypothesis is on the task list 
of modern experiments.  %Please define CPT. 

Despite the great successes of the SM, we can hardly believe that it is the true fundamental
theory of Nature. Most likely, it is an effective model with a limited applicability domain.
The search for the upper energy limit of the SM applicability is the actual task at
all high-energy colliders experiments. Up to now, all direct attempts to find elementary
particles and interactions beyond the Standard Model have failed. The accent of experimental
studies has shifted towards accurate verification of the SM features.
Deep investigation of the SM symmetries is an important tool in this line of research. 

Asymmetries form a special class of experimental observables. First of all, they explicitly
access the breaking of a certain symmetry in Nature. Second, they are usually constructed
as a ratio of observed quantities, in which the bulk of experimental and theoretical
systematic uncertainties is canceled out. So the asymmetries provide independent
additional information on particle interactions. They are especially sensitive
to non-standard weak interactions including contributions of right currents and
new intermediate $Z^\prime$ vector bosons, see e.g.,~\cite{Fujii:2019zll}. 

The physical programs of future (super) high-energy electron--positron colliders such
as CLIC~\cite{Aicheler:2012bya}, ILC~\cite{MoortgatPick:2005cw,Baer:2013cma,Bambade:2019fyw}, FCC-ee~\cite{Abada:2019zxq},
and CEPC~\cite{CEPC-SPPCStudyGroup:2015csa} necessarily include accurate tests of the SM.
Studies of polarization effects and asymmetries will be important to probe of the fundamental properties of Higgs boson(s) and, in particular, in the process of annihilation into 
top quarks~\cite{BhupalDev:2007ftb,Hagiwara:2016rdv,Ma:2018ott}.
The future colliders plan to start operation in the so-called GigaZ mode at the $Z$ peak
and improve upon the LEP both in statistical and systematical uncertainties in tests of
the SM~\cite{Erler:2000jg} by at least one order of magnitude. 
Among these collider projects, the FCC-ee one has the most advanced program of high-precision
measurements of SM processes at the $Z$ peak.
Such tests have been performed at LEP and SLC and they have confirmed the validity of the SM
at the electroweak (EW) energy scale of about 100~GeV~\cite{Grunewald:1999wn,ALEPH:2005ab}.  
During the LEP era, extensive experimental and theoretical studies of asymmetries
made an important contribution to the overall verification of the SM, see review~\cite{Mnich:1996hy}
and references therein. %Please define LEP, CLIC, ILC, FCC, CEPC. Please ensure all terms/acronyms/abbreviations are defined upon first usage (1) in the abstract, (2) in the text, and (3) in a figure or in a table. So, terms may need to be defined up to three times in a manuscript.
The new precision level of future experiments motivates us to revisit the
asymmetries and scrutinize the effects of radiative corrections (RCs) to them.
In the analysis of LEP data, semi-analytic computer codes like {\tt ZFITTER}~\cite{Arbuzov:2005ma}
and {\tt TOPAZ0}~\cite{Montagna:1998kp} were extensively used. The forthcoming new generation
of experiment requires more advanced programs, primarily Monte Carlo event generators.

The article is organized as follows. The next section contains preliminary remarks 
and the general notations. Section~\ref{Sect:ALR} is devoted to the left--right asymmetry.
The forward--backward asymmetry is considered in Section~\ref{Sect:AFB}. Discussion of 
the left--right forward--backward asymmetry is presented in Section~\ref{Sect:Alrfb}. 
In Section~\ref{Sect:Ptau}, we provide results related to the final state fermion polarization. 
Section~\ref{Sect:Concl} contains a discussion and conclusions.

\section{Preliminaries and Notations}

In the recent paper~\cite{Bondarenko:2020hhn} by the SANC group,
high-precision theoretical predictions 
for the process $e^+e^- \to l^+ l^-$ ($l=\mu$ or $\tau$) were presented.
With the help of computer system {\tt SANC}~\cite{Andonov:2004hi},
we calculated the complete 1-loop electroweak radiative
corrections to these processes, taking into account possible longitudinal polarization
of the initial beams. The calculations were performed within the 
helicity amplitude formalism, taking into account the initial and final
state fermion masses. So, the {\tt SANC} system provides a solid framework 
to access asymmetries in $e^+e^-$ annihilation processes and to study various 
relevant effects. In particular, the system allows us to separate effects due to
quantum electrodynamics (QED) and weak radiative corrections. %Please define QED. 

The focus of this article is on the description and assessment of the
asymmetry family: the~left--right asymmetry $\alr$, 
the forward--backward asymmetry $\afb$, the left--right forward--backward  asymmetry $\alrfb$,
and the final state fermion polarization $\ptau$ 
in collisions of high-energy polarized or unpolarized $e^+e^-$ beams. 
The main aim was to verify the effect of radiative corrections on 
the extraction of the SM parameters from the asymmetries and to analyze 
the corresponding theoretical~uncertainty.

We performed calculations for polarized initial and final state particles.
Beam polarizations play an important role: 
\begin{itemize}
\item They improve the sensitivity
to CP-violating anomalous couplings or form factors, which
are measurable even with unpolarized beams through the forward--backward asymmetry. %please define CP. 
\item With the polarization of both beams, the sensitivity to the
new physics scale can be increased by a factor of up to 1.3 with respect
to the case with only polarized electrons~\cite{Fujii:2019zll}.
\item A high-luminosity at the GigaZ stage of a collider running at 
the $Z$ boson resonance with positron polarization allows us to improve 
the accuracy of the determination of $\sin^2\vartheta_W$ ($\vartheta_W$
is the electroweak mixing angle) by an order of magnitude,
through studies of the left--right asymmetry~\cite{Fujii:2019zll}.
\end{itemize}

Numerical illustrations for each asymmetry are given in two
energy domains: the wide center-of-mass energy range $20\leq \sqrt{s} \leq 500$~GeV
and the narrow one around the $Z$ resonance ($70\leq \sqrt{s} \leq 100$~GeV), where
a peculiar behavior of observables can be seen.
All results were produced with the help of the $e^+e^-$ branch~\cite{Arbuzov:2019pxw} 
of the {\tt MCSANC} Monte Carlo integrator~\cite{Arbuzov:2015yja}.

Let us introduce the notation.
First of all, we define quantities $A_f$ $(f=e,\mu,\tau)$ which are often used
for description of asymmetries at the $Z$ peak:
\bqa
A_f \equiv 2 \frac{g_{{\sss V}_f}g_{{\sss A}_f}}{g^2_{{\sss V}_f}+g^2_{{\sss A}_f}}
=\frac{1-(g_{R_f}/g_{L_f})^2}{1+(g^2_{R_f}/g^2_{L_f})^2},
\eqa
where the vector and axial-vector coupling constants of the weak neutral current
of the fermion $f$ with the electromagnetic charge $q_f$
(in the units of the positron charge $e$) are 
\bqa
g_{{\sss V}_f}\equiv I^3_f- 2 q_f \sin^2\vartheta_W, \qquad
g_{{\sss A}_f}\equiv I^3_f.
\eqa

The corresponding left and right fermion couplings are
\bqa
g_{L_f}\equiv I^3_f- q_f \sin^2\vartheta_W, \qquad
g_{R_f}\equiv - q_f \sin^2\vartheta_W.
\eqa

The neutral current couplings $g_{L_f}$ and $g_{R_f}$ quantify the strength
of the interaction between the $Z$ boson and the given chiral states of the fermion.

We claim that there are sizable corrections to all observable asymmetries
due to radiative corrections which affect simple Born-level analytic 
formulae relating the asymmetries with electroweak parameters.
It is especially interesting to consider the
behavior of asymmetries in different EW schemes: $\alpha(0)$, $\alpha(\mz^2)$, 
and $G_\mu$, see their definitions below. 
We also will compare the results in the Born and 1-loop approximation. 
The latter means inclusion of 1-loop radiative corrections of one
of the following types: pure QED photonic RCs (marked as ``QED''), 
weak RCs (marked as ``weak''), and the complete 1-loop electroweak RCs
(marked as ``EW''): 
$$\sigma_{\mathrm{EW}} = \sigma_{\mathrm{Born}} + \sigma_{\mathrm{QED}} + \sigma_{\mathrm{weak}}.$$

The weak part in our notation includes 1-loop self-energy corrections to photon 
and $Z$ boson propagators. In our notation, higher-order effects due to interference 
of pure QED and weak contributions are a part of $\sigma_{\mathrm{weak}}$.

The cross section of a generic annihilation process
of longitudinally polarized $e^+$ and $e^-$ with polarization degrees
$P_{e^+}$ and $P_{e^-}$ can be expressed as follows:
\bqa
\sigma(P_{e^-},P_{e^+}) &=& (1+P_{e^-})(1+P_{e^+}) {\sigma}_{RR}
 +(1-P_{e^-})(1+P_{e^+}) {\sigma}_{LR}
\nll
 &+& (1+P_{e^-})(1-P_{e^+}) {\sigma}_{RL}
 +(1-P_{e^-})(1-P_{e^+}) {\sigma}_{LL}.
\label{PolXSec}
\eqa

Here $\sigma_{ab} = \sum_{ij(k)}\lvert{\cal H}_{abij(k)}\rvert^2$ are the $2 \to 2 (3)$
helicity amplitudes of the reaction, $(ab=RR,RL,LR,LL)$ with right-handed $R$=''$+$''
or left-handed $L$=``$-$'' initial particles.

It is convenient to combine the electron $P_{e^-}$ and positron $P_{e^+}$ 
polarizations into the effective~quantity
\bqa
P_{\mathrm{eff}} = \frac{P_{e^-}-P_{e^+}}{1-P_{e^-}P_{e^+}}.
\eqa

In the case when only the electron beam is polarized, the effective polarization 
coincides with the electron one.

To investigate theoretical uncertainties, we use the following three EW schemes:
\begin{enumerate}
\item the $\alpha(0)$ scheme in which the fine-structure constant $\alpha(0)$ is used
as input. The contribution of RCs in this scheme is enhanced by the large logarithms
of light fermion masses via $\alpha(0) \ln(s/m_f^2)$~terms.
\item The $\alpha(\mz^2)$ scheme in which the effective electromagnetic constant
$\alpha(\mz^2)$ is used at Born level while virtual 1-loop and real photon
bremsstrahlung contributions are proportional to $\alpha^2(\mz^2)\alpha(0)$.
In this scheme the virtual RCs receive contributions from the quantity $\Delta\alpha(\mz^2)$
which describes the evolution of the electromagnetic coupling from the scale $Q^2=0$ to 
the $Q^2=\mz^2$ one and cancels the large terms with logarithms of light fermion masses.
\item the $G_{\mu}$ scheme in which the Fermi coupling constant $G_{\mu}$,
extracted from the muon life time, is used
at the Born level while the virtual 1-loop and real photon
bremsstrahlung contributions are proportional to $G_{\mu}^2\alpha(0)$.
The virtual RCs receive contributions from the quantity $\Delta r$.
Since the expression for $\Delta r$ contains the $\Delta\alpha(\mz^2)$,
the large terms with logarithms of the light masses are also canceled.
The quantity $\Delta r$ rules the $G_{\mu}$ and $\alpha(0)$
relation in this scheme.
\end{enumerate}

Results of fixed-order perturbative calculations in these schemes differ 
due to missing higher-order effects.
In what follows, numerical calculations are performed in the $\alpha(0)$
EW scheme if another choice is not explicitly indicated.

\section{Left--Right Asymmetry \boldmath{$\alr$}} \label{Sect:ALR}

A scheme to measure the $\alr$ polarization asymmetry at the $Z$ peak was
suggested in ~\cite{Blondel:1987wr}. It~was shown that this observable
can be used as for extraction of electroweak couplings as well as for
a polarimeter calibration.

If we neglect the initial electron masses, the polarized cross-section 
can be rewritten in the following form:
\bqa
\sigma(P_{e^-},P_{e^+}) &=& (1-P_{e^-} P_{e^+}) [ 1 - P_{\mathrm{eff}} \alr] {\sigma}_{0},
\label{PolXSecLR}
\eqa
where ${\sigma}_{0}$ is the unpolarized cross-section.

The left--right asymmetry in the presence of partially polarized ($|P_{\mathrm{eff}}|<1$)
initial beams is defined~as
\bqa \label{alr_def}
A_{\rm LR} = \frac{1}{P_{\mathrm{eff}}}
\frac{\sigma(-P_{\mathrm{eff}})-\sigma(P_{\mathrm{eff}})}
{\sigma(-P_{\mathrm{eff}})+\sigma(P_{\mathrm{eff}})},
\label{ALRpol}
\eqa
where $\sigma$ is the cross-section with polarization $P_{\mathrm{eff}}$.

In the case of fully polarized initial particles ($|P_{e^\pm}|=1$) the definition (\ref{ALRpol}) becomes:
%The left--right asymmetry is defined as
\bqa
A_{\rm LR} = \frac{\sigma_{L_{e}}-\sigma_{R_{e}}}{\sigma_{L_e}+\sigma_{R_e}},
\eqa
where $L_e$ and $R_e$ refer to the left and right helicity
states of the incoming electron.

Equations~(\ref{PolXSecLR}) and (\ref{ALRpol}) show that $\alr$
does not depend on the degree of the initial beam polarization.  

This type of asymmetry is sensitive to weak interaction effects in the initial vertex.
In the Born approximation at energies close to the $Z$ resonance, 
it is directly related to the electron coupling:
\bqa
A_{\rm LR} \approx A_e.
\label{ALRz}
\eqa

\begin{figure}[h]
\centering
\includegraphics[width=0.45\linewidth]{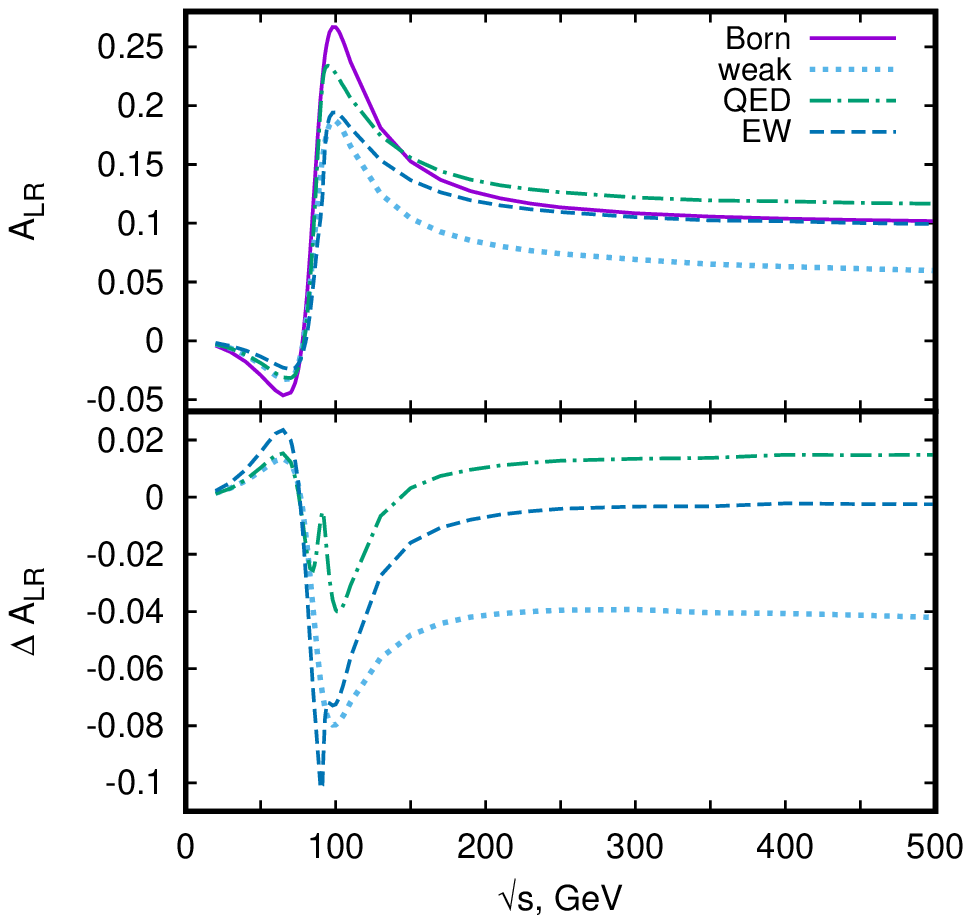}
\includegraphics[width=0.45\linewidth]{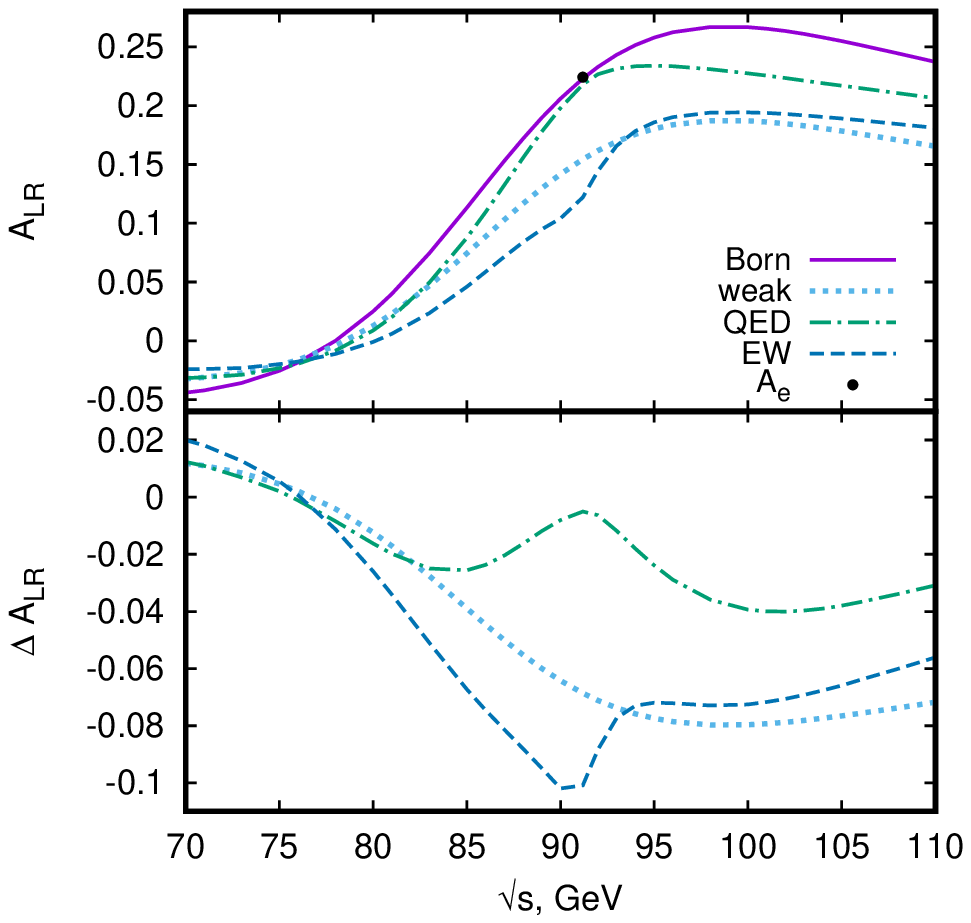}
\caption{\label{fig-alr}
  (\textbf{Left}) The $\alr$ asymmetry in the Born and 1-loop (weak, pure quantum electrodynamics (QED), and electroweak (EW)) approximations
  and $\Delta\alr$ vs. center-of-mass system (c.m.s.) energy in a wide range;
  (\textbf{Right}) the same for the $Z$ peak region.
}
\end{figure}

The left--right asymmetry $A_{\rm LR}$ as a function of the center-of-mass system (c.m.s.) 
energy in the ranges $20\leq \sqrt{s}\leq 500$~GeV (Left) and 
$70\leq \sqrt{s}\leq 110$~GeV (Right) 
is shown in Figure~\ref{fig-alr}. We~explore $\alr$ in different approximations
and the corresponding shifts $\Delta\alr$ between the Born level
and 1-loop corrected approximations taking into account 
either pure QED, or weak, or complete EW effects: 
$\Delta\alr$=$\alr$(1-loop corrected)-$\alr$(Born).
The right figure shows the behavior of $\alr$ near the $Z$ resonance,
and the value $A_e$ at $\sqrt{s} =M{\sss Z}$ is indicated by a black dot (see (\ref{ALRz})).

One can notice that although the total 1-loop EW corrections to the process 
cross-section are equal to the sum of the pure QED and weak ones,
the corresponding shifts $\Delta\alr$ are not additive. 
That is because the asymmetry is defined as a ratio and the corrections
affect both the numerator and~denominator. 

\begin{figure}[h]
\centering
\includegraphics[width=0.5\linewidth]{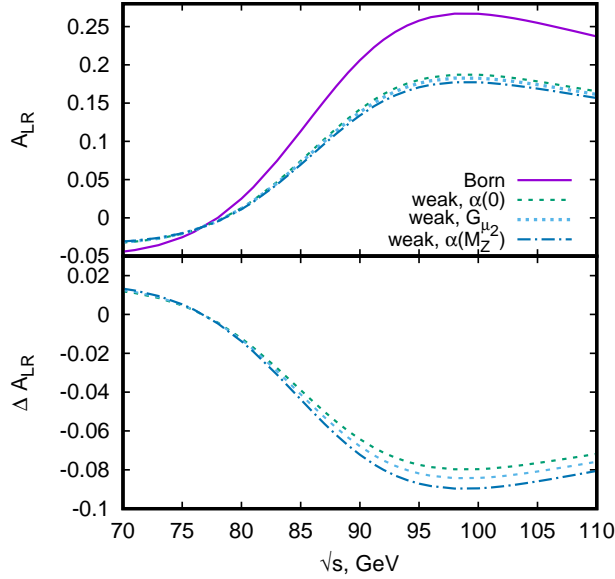}
\caption{\label{fig-alr-024}
%  (Left)
  The $\alr$ asymmetry at the Born level and with 1-loop weak radiative corrections (RCs); the corresponding
  shifts $\Delta\alr$ within $\alpha(0)$, $G_{\mu}$, and $\alpha(\mz^2)$ EW schemes
  vs. c.m.s. energy in the peak~region.
}
\end{figure}

In Figure~\ref{fig-alr-024} we show $\alr$ for the Born and weak 1-loop corrected 
levels of accuracy in different EW schemes and the corresponding shifts
$\Delta\alr$=$\alr$(weak, some EW scheme)-$\alr$(Born).
We see that the effects due to weak corrections in different EW schemes
behave in a similar way. Nevertheless the scheme dependence is visible within
the expected precision of future measurements. The deviations between the results
in different schemes can be treated as a contribution into the theoretical uncertainty
due to missing higher order corrections. 

\unskip

The impact of 1-loop EW contributions to $\Delta\alr$ is of the order
$-0.1$ in the resonance region, but at energies above $\sqrt{s}=200$~GeV
there are considerable cancellations between weak and QED effects so 
that the combined EW corrections becomes small (but still numerically
important for high-precision measurements). 

{\bf Summary for $\alr$}

The left--right asymmetry $\alr$ is almost insensitive to the details of 
particle detection since the corresponding experimental uncertainties
tend to cancel out in the ratio~(\ref{alr_def}). 
It (almost) does not depend on the final state fermion couplings in the vicinity
of the $Z$ boson peak and can be measured for 
any final state with a large gain in statistics.
For this reasons it is appropriate for extraction of the 
$\sin^2\vartheta_W^{\mathrm{eff}}$ value.

We observe that the values $\Delta\alr$ due to weak and pure QED 1-loop corrections
are very significant at high energies in general, but in the resonance region
impact of QED is small, while the weak contribution to $\Delta\alr$ reaches 0.07.
Therefore, it is necessary to evaluate all possible radiative correction contributions 
to the weak parts of RCs carefully and thoroughly.

\section{Forward--Backward Asymmetry \boldmath{$\afb$}} \label{Sect:AFB}

The forward--backward asymmetry is defined as
\bqa
&& A_{\rm FB} = \frac{\sigma_{\rm F}-\sigma_{\rm B}}{\sigma_{\rm F}+\sigma_{\rm B}},
\nonumber \\
&& \sigma_{\rm F} = \int\limits_0^1 \frac{d\sigma}{d\cos\vartheta_f}d\cos\vartheta_f,
\qquad
\sigma_{\rm B} = \int\limits_{-1}^0 \frac{d\sigma}{d\cos\vartheta_f}d\cos\vartheta_f,
\eqa
where $\vartheta_f$ is the angle between the momenta of the incoming electron and the outgoing
negatively charged fermion.
It can be measured in any $e^+e^- \to f \bar{f}$ channels but
for precision test the most convenient channels are $f=e,\mu$.
The channels with production of $\tau$ leptons, $b$ or $c$ quarks 
are very interesting as well.

At the Born level, this asymmetry is proportional to the product of initial
and final state couplings and is caused by parity violation at both production
and decay vertices:
\bqa
A_{\rm FB} \approx \frac{3}{4} A_e A_f.
\label{AFBz}
\eqa

In the case of partially polarized initial beams the condition (\ref{AFBz}) reduces
to the following one
\bqa
A_{\rm FB} \approx \frac{3}{4} \frac{A_e - P_{\rm eff}}{1-A_e P_{\rm eff}} A_f.
\label{AFBzpol}
\eqa

\begin{figure}[h]
\centering
\includegraphics[width=0.45\linewidth]{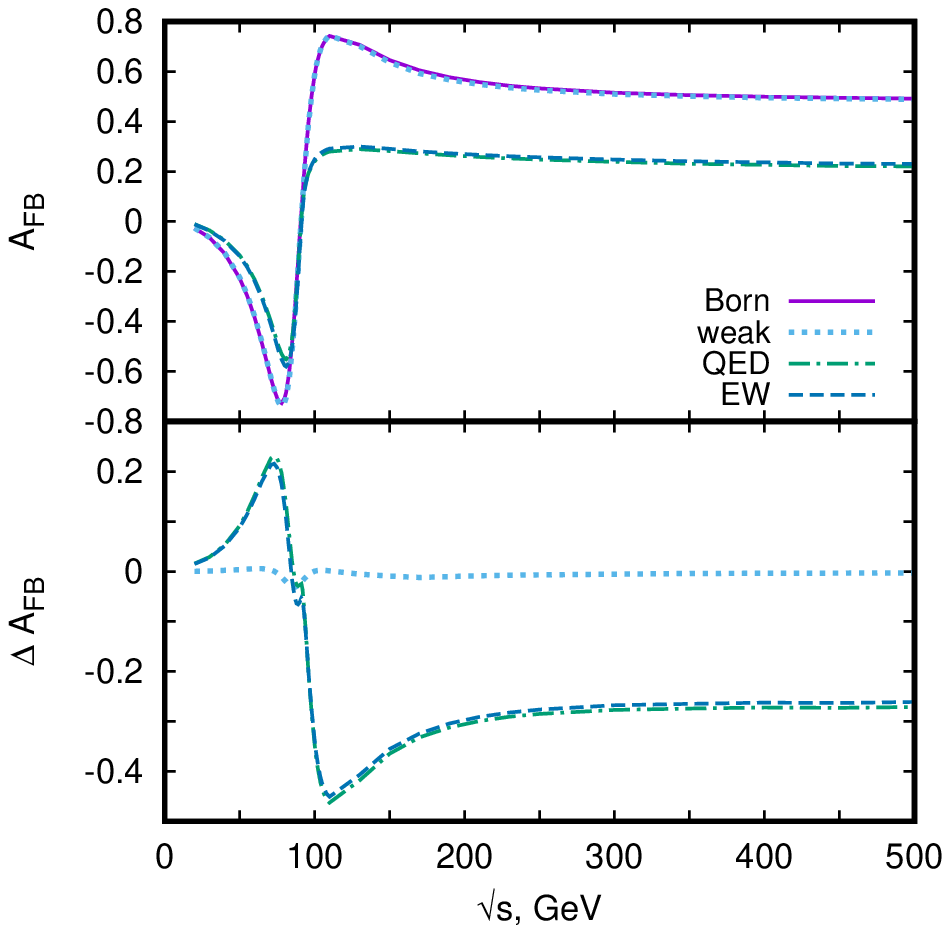}
\includegraphics[width=0.45\linewidth]{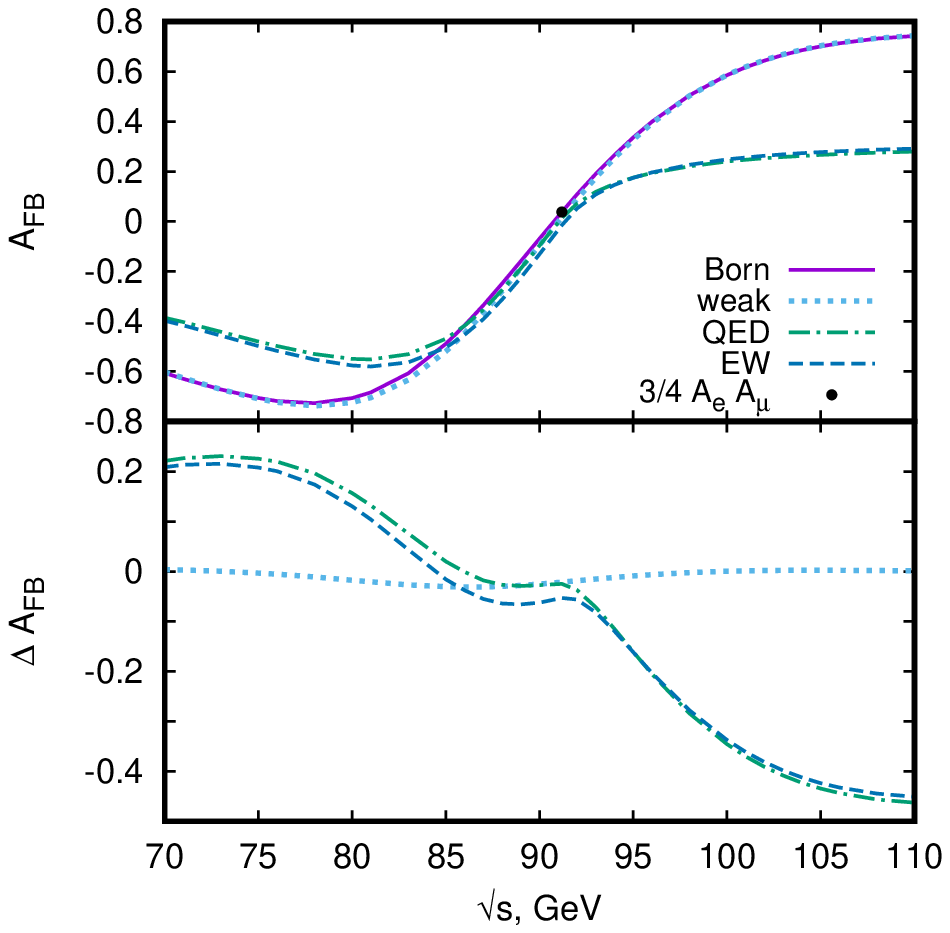}
\caption{\label{fig-afb}
  (\textbf{Left}) The $\afb$ asymmetry in the Born and 1-loop (weak, QED, EW) approximations
  and the corresponding shifts $\Delta\afb$ for a wide c.m.s. energy range;
  (\textbf{Right}) the same for the $Z$ peak region.
}
\end{figure}

In Figure~\ref{fig-afb} we show the behavior of the $\afb$ asymmetry
in the Born and 1-loop approximations (with weak, pure QED, or complete
EW contributions) 
and the corresponding $\Delta\afb$ for c.m.s. energy range $20\leq\sqrt{s}\leq 500$~GeV 
in the left plot and for the $Z$ peak region of c.m.s. energy $70\leq\sqrt{s}\leq 110$~GeV 
in the right one. As in the previous case of $\alr$, we 
indicate by a black dot the value $\afb \approx 3/4 A_e A_\mu $ at the resonance. 
We observe that the weak contribution to $\afb$ is small and practically does not depend  
on energy. The shift $\Delta\afb$ changes the sign at the resonance and tends to 
a constant value $(\sim$$-0.3)$ above $200$~GeV.
The huge magnitude of the shift $\Delta\afb$ out of the $Z$ resonance region 
is coming mainly from the pure QED corrections. In particular, above the peak 
the effect due to radiative return to the resonance is very important.

\begin{figure}[h]
\centering
  \includegraphics[width=0.45\linewidth]{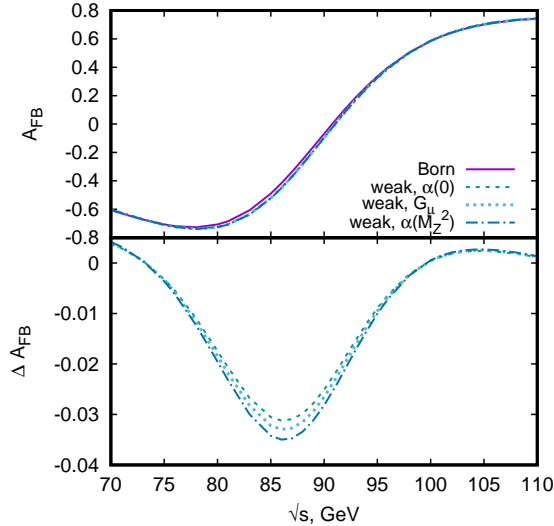}
\caption{\label{fig-afb-024}
  The $\afb$ asymmetry and $\Delta\afb$ in the Born and complete 1-loop EW approximations
  within the $\alpha(0)$, G$_{\mu}$, and $\alpha(\mz^2)$ EW schemes vs. the
  c.m.s energy.}
\end{figure}

Figure~\ref{fig-afb-024} shows the dependence of $\afb$ for different
levels of accuracy (Born and 1-loop weak) on the EW scheme choice:
either $\alpha(0)$, or G$_{\mu}$, or $\alpha(\mz^2)$.
The corresponding shifts $\Delta\afb$ between
the Born and the 1-loop weak corrected approximations
are shown in the lower plot.

Below we investigate two sets of polarization degree $P_i=(P_{e^-},P_{e^+})$:
\bqa
P_1 = (-0.8,0.3) \qquad \mathrm{and} \qquad P_2 = (0.8,-0.3).
\label{p12}
\eqa

\begin{figure}[h]
\centering
  \includegraphics[width=0.45\linewidth]{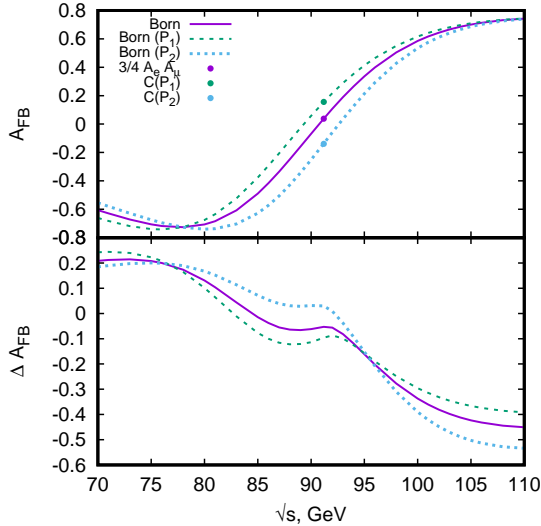}
\caption{ \label{fig:afb_p12}
  The $\afb$ asymmetry at the Born level (upper panel) and the corresponding $\Delta\afb$ 
  in the 1-loop EW 
  approximation (bottom panel) for unpolarized and polarized cases with degrees of beam 
  polarizations $P_{1,2}$~(\ref{p12})  
  vs. c.m.s. energy in the $Z$ peak region. The constants $C(P_{1,2})$ stand for the expression
  (\ref{AFBzpol}) with polarization degrees~(\ref{p12}).
}
\end{figure}

In Figure~\ref{fig:afb_p12} we compare the values of $\afb$ asymmetry
and the corresponding shifts due to EW corrections for the unpolarized
case and two choices of polarized beams defined in the above equation.
One can see that a combination of polarization degrees of initial particles 
can either increase or decrease 
the magnitude of the $\afb$ asymmetry with respect to the unpolarized case. 

There is an interesting idea~\cite{Janot:2015gjr} to use the $A_{\rm FB}$ asymmetry
at the FCC-ee in order to directly access the value of QED running coupling 
at $\mz$. This idea was supported in ~\cite{Jadach:2018lwm} where it was 
demonstrated that higher-order QED radiative corrections to $A_{\rm FB}$ are under control.
Our results show that higher-order effects due to weak interactions are not negligible 
in this observable; further studies are required. 

At the Born level there are contributions suppressed by the small factor $m_f^2/s$ 
with the fermion mass squared. It is interesting to note that in 1-loop radiative corrections
there are contributions of the relative order $\alpha\cdot m_f/\sqrt{s}$ with the fermion
mass to the first power~\cite{Arbuzov:1991pr}, which are numerically relevant at high energies
especially for the $b$ quark channel.

{\bf Summary for $\afb$}

The weak 1-loop contribution $\Delta\afb$ is rather small for the whole energy 
range, see Figure~\ref{fig-afb}. Nevertheless in this
asymmetry the difference between the pure QED and the complete 1-loop approximations
near the resonance is numerically important. The dependence on the EW scheme choice, 
see Figure~\ref{fig-afb-024}, is small but still relevant for high-precision measurements.
The dependence of this asymmetry on polarization is very significant.

\section{Left--Right Forward--Backward  Asymmetry \boldmath{$\alrfb$}} \label{Sect:Alrfb}

In order to measure the weak couplings of the final state fermions,
it was suggested to analyze the so-called left--right forward--backward
asymmetry~\cite{Blondel:1987gp}:
%\bqa
%A^{pol}_{FB} =\frac{\sigma(f_L)-\sigma(f_R)}{\sigma(f_L)+\sigma(f_R)}.
%\eqa
%
%\bqa
%A_{\rm lrfb} =
%\frac{\left[\sigma(f_{L})-\sigma(f_{R})\right]_{F}-\left[\sigma_{L}-\sigma_{R}\right]_{B}}
%     {\left[\sigma(f_{L})+\sigma(f_{R})\right]_{F}+\left[\sigma_{L}+\sigma_{R}\right]_{B}}
%\eqa
\bqa
\alrfb =
\frac{\left(\sigma_{L_e}-\sigma_{R_e}\right)_{F}-\left(\sigma_{L_e}-\sigma_{R_e}\right)_{B}}
     {\left(\sigma_{L_e}+\sigma_{R_e}\right)_{F}+\left(\sigma_{L_e}+\sigma_{R_e}\right)_{B}},
\label{alrfb}
\eqa
where $\sigma_{L}$ and $\sigma_{R}$ are the cross sections with
left and right handed helicities of the initial electrons.

From the definition (\ref{alrfb}) it follows that $\alrfb$ partially inherits the properties of the $\alr$ and, in~particular, 
does not depend on the degree of the initial beam polarizations.

%\bqa
%$\alrfb$
%=\frac{(\sigma_{L}-\sigma_{R})_F-(\sigma_{L}-\sigma_{R})_B}
%      {(\sigma_{L}+\sigma_{R})_F+(\sigma_{L}+\sigma_{R})_B}.
%\eqa

In the case of unpolarized beams on the $Z$ resonance peak, the Born-level asymmetry 
is
\bqa \label{alrfb_approx}
\alrfb \approx \frac{3}{4} A_f.
\eqa

\begin{figure}[h]
\centering
\includegraphics[width=0.45\linewidth]{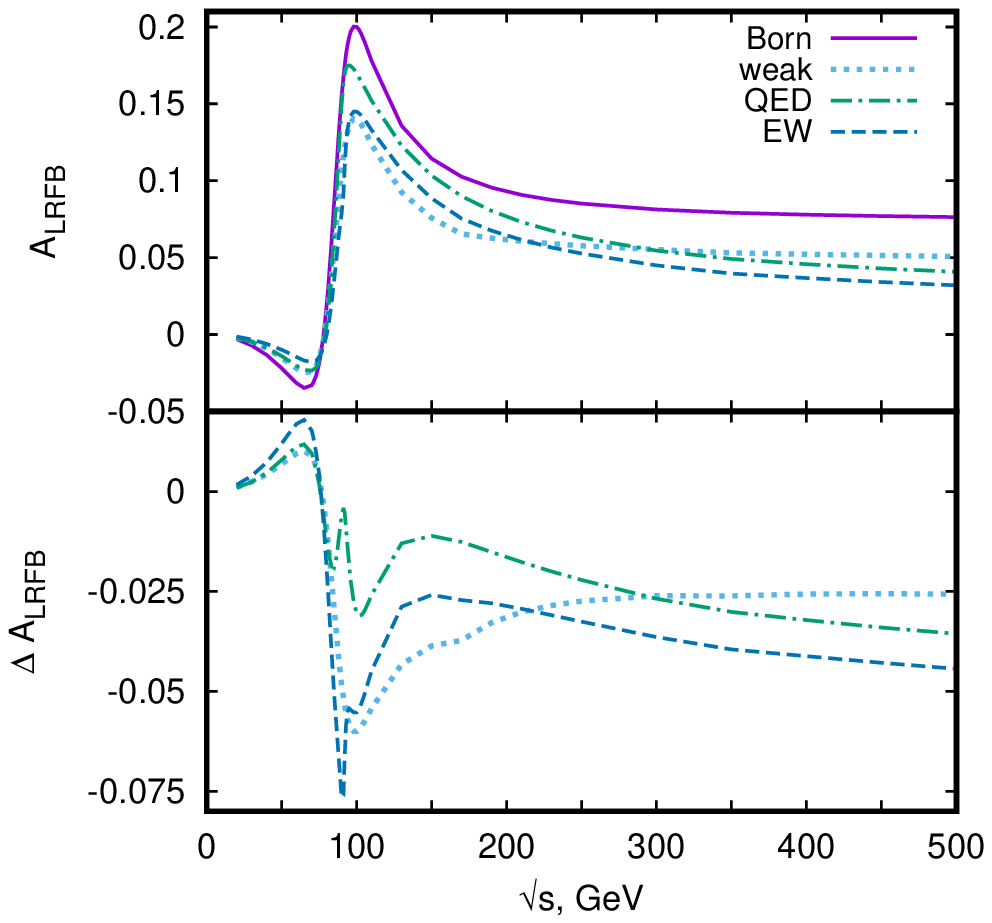}
\includegraphics[width=0.45\linewidth]{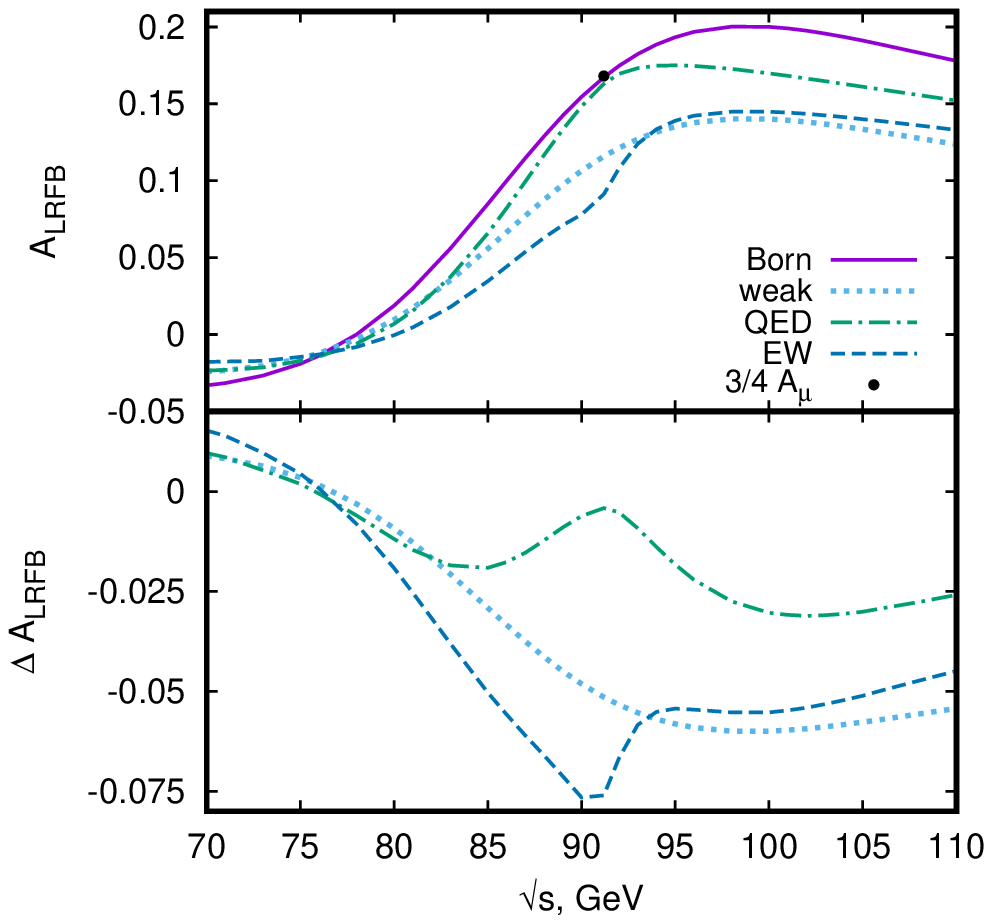}
\caption{ \label{fig_alrfb}
  (\textbf{Left}) The $\alrfb$ asymmetry in the Born and 1-loop (weak, QED, EW) approximations 
  and $\Delta\alrfb$ for c.m.s. energy  range;
  (\textbf{Right}) the same for the $Z$ peak region.
}
\end{figure}

In Figure~\ref{fig_alrfb} we present the predictions for the $\alrfb$ asymmetry
in several approximations, namely at the Born level and with 1-loop weak, pure QED, and complete EW contributions.

\begin{figure}[h]
\centering
\includegraphics[width=0.45\linewidth]{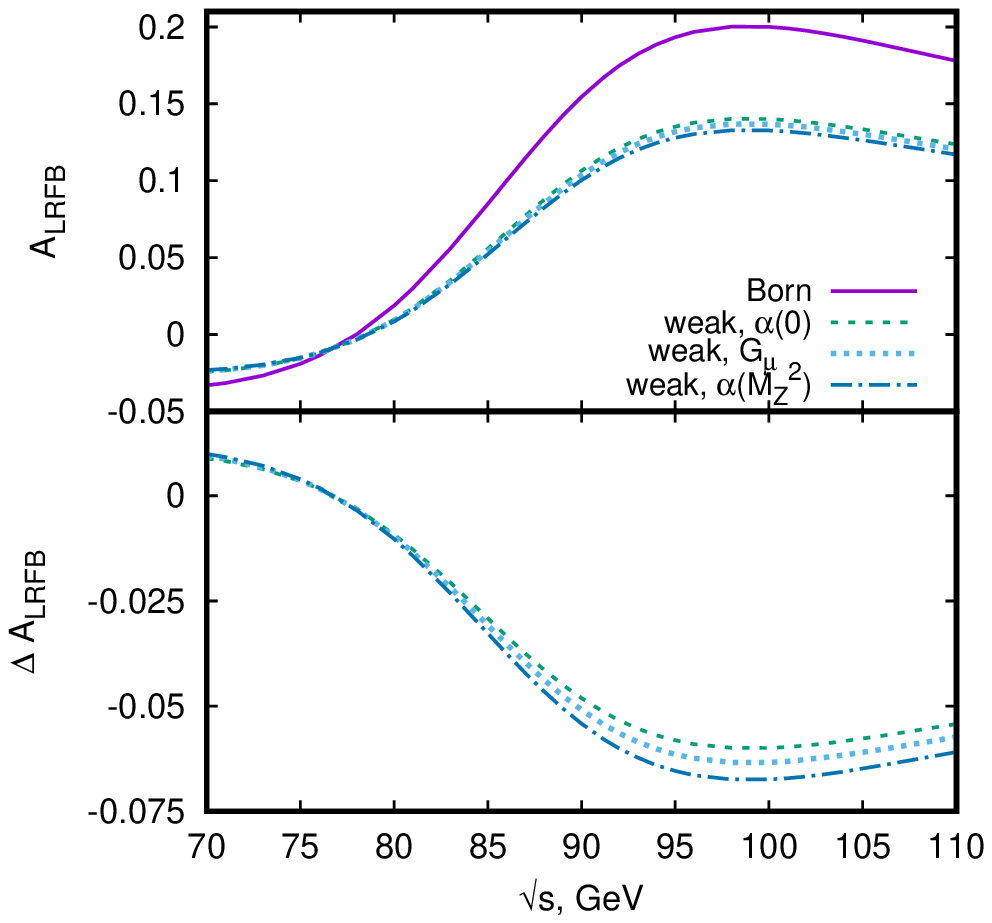}
\caption{\label{fig-alrfb}
%  (Left)
  The $\alrfb$ asymmetry in the Born and 1-loop EW approximations
  and $\Delta\alrfb$ within $\alpha(0)$, G$_{\mu}$, and $\alpha(\mz^2)$ EW schemes
  vs. c.m.s. energy in the $Z$ peak region.
}
\end{figure}

Next, we repeat the study of the $\alrfb$ asymmetry behavior in different EW schemes.
We have illustrated the energy dependence of the $\alrfb$ asymmetry in
$\alpha(0)$, G$_{\mu},$ and $\alpha(\mz^2)$ schemes
and the corresponding $\Delta\alrfb$ in Figure~\ref{fig-alrfb}.
The impact of weak corrections on $\alrfb$ is large. For example, the Born-level
value of $\alrfb$ at the $Z$ peak is about $0.17$, while accounting for the weak RCs
contribution reduces the asymmetry value down to $\sim$0.11.

{\bf Summary for $\alrfb$}

We would like to emphasize that the above Formula~(\ref{alrfb_approx}) 
appears to be a rather rough approximation since radiative corrections shift
the observable value of $\alrfb$ quite a lot.
Apparently the $\alrfb$ asymmetry is more affected by weak corrections than $\alr$.
The shifts $\Delta\alrfb$ only slightly depend on an EW scheme choice.
The $\alrfb$ asymmetry at the $Z$ boson peak depends on the final lepton coupling that could be used 
to measure the $\mu$ and $\tau$ weak couplings and their difference from the initial lepton (electron) one.

\section{Final-State Fermion Polarization \boldmath{$P_f$}}  \label{Sect:Ptau}

The polarization of a final-state fermion $P_{f=\mu,\tau}$ can be expressed as 
the ratio between the difference of the cross sections for right and left handed 
final state helicities and their sum
\bqa
P_f =\frac{\sigma_{R_f}-\sigma_{L_f}}{\sigma_{R_f}+\sigma_{L_f}}.
\eqa

In an experiment, it can be measured for the $\tau^+\tau^-$ channel by reconstructing 
the $\tau$ polarization from the pion spectrum in the decay $\tau \to \pi\nu$.
Details of the analysis of $\ptau$ measurements at LEP are 
described in~\cite{ALEPH:2005ab}. Computer programs {\tt TAOLA}~\cite{Jadach:1993hs} 
and {\tt KORALZ}~\cite{Jadach:1993yv,Jadach:1999tr} were applied for this analysis.
Estimated improvement for $\ptau$ and $\tau$ decay products over LEP time  
in  ILC  in the GigaZ program was done in \cite{Bambade:2019fyw}.

In the case for unpolarized beams in the vicinity of the $Z$ peak, the expression for
channel $e^+e^- \to \tau^+\tau^-$  is simplified to
\bqa
\ptau(\cos\vartheta_{\tau}) \approx -\frac{\ds A_\tau   + \frac{2 \cos\vartheta_{\tau}}{1+\cos^2\vartheta_{\tau}}A_e}
    {\ds 1 + \frac{2 \cos\vartheta_{\tau}}{1+\cos^2\vartheta_{\tau}}A_e A_\tau}.
\eqa

From this observable, one can extract information on the couplings $A_\tau$ and $A_e$, simultaneously.

\begin{figure}[h]
\centering
  \includegraphics[width=0.45\linewidth]{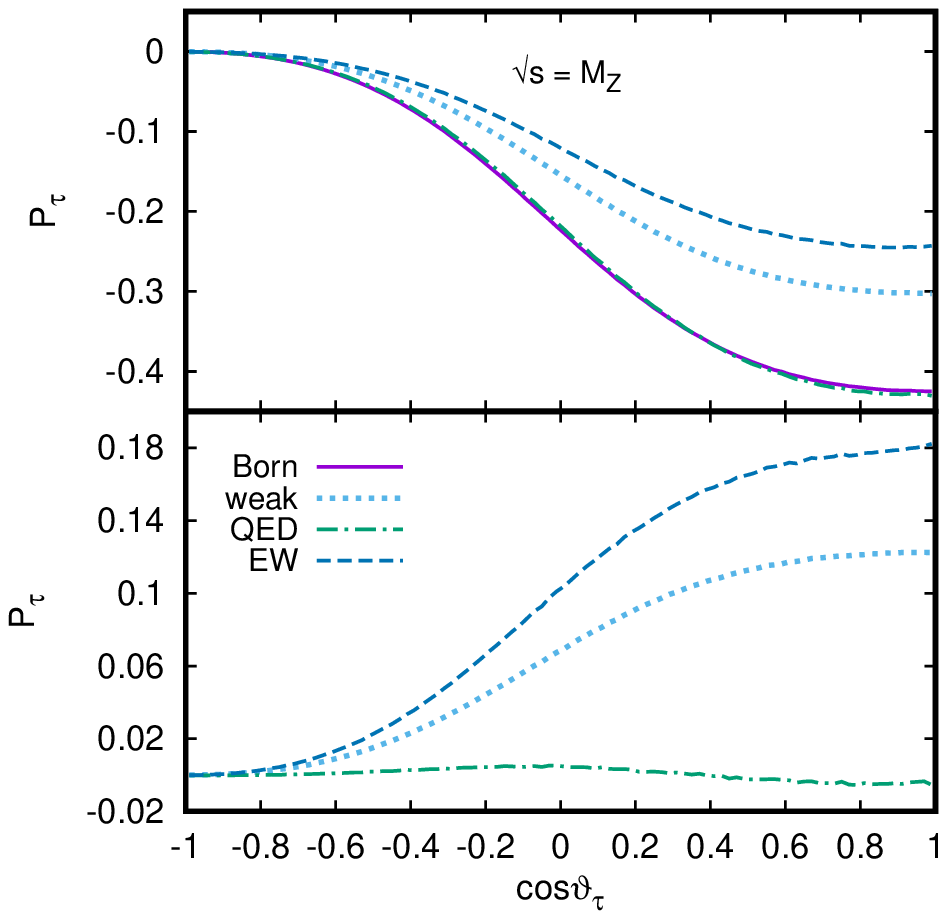}
  \includegraphics[width=0.45\linewidth]{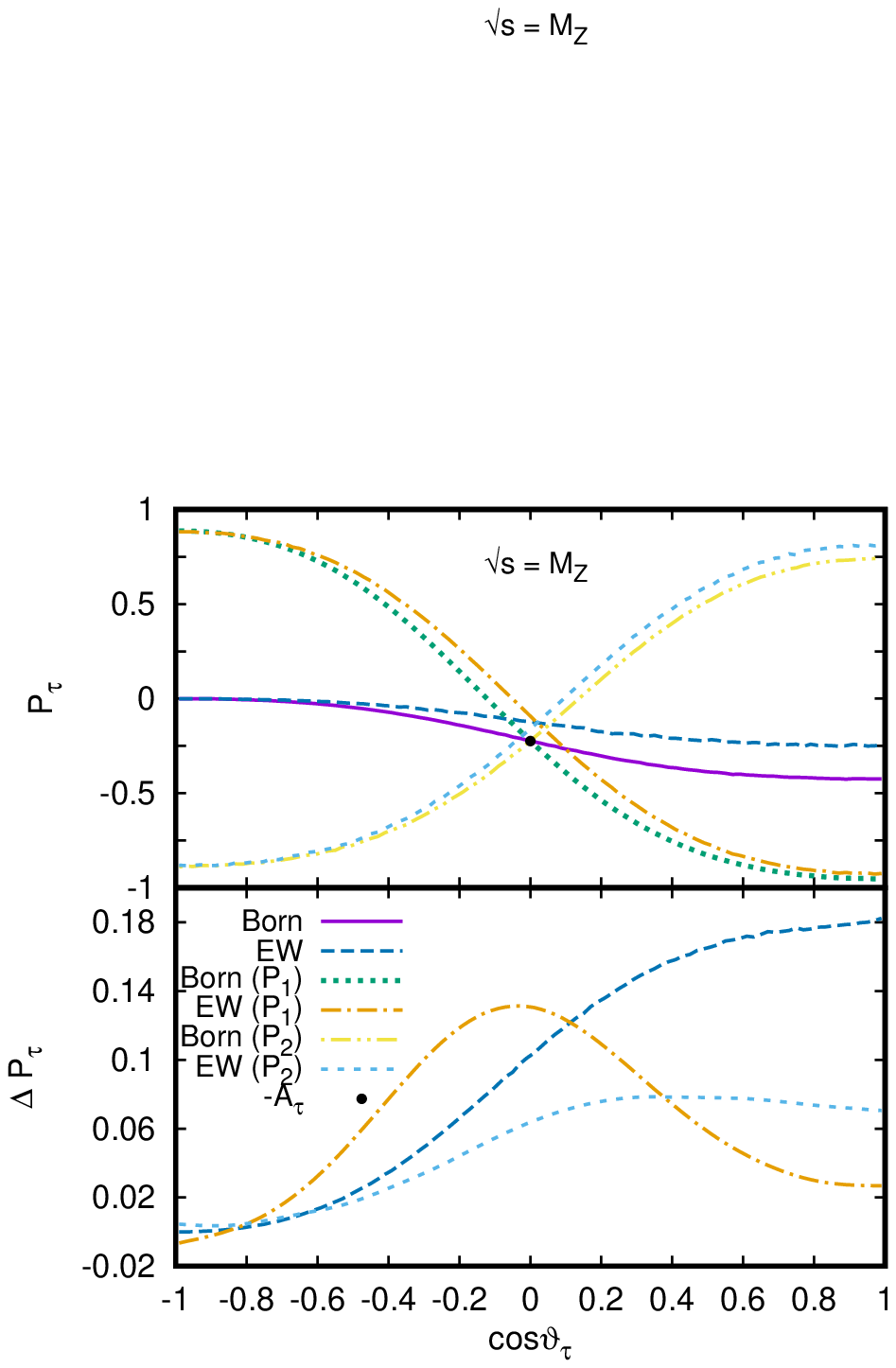}
  \caption{\label{ptau_B_w_QED_EW}
(\textbf{Left}) The $\ptau$ polarization in the Born and 1-loop (weak, pure QED, and EW) approximations
    as a function of $\cos{\vartheta_{\tau}}$ at $\sqrt{s}=\mz$.
(\textbf{Right}) The $\ptau$ polarization for unpolarized and polarized cases with (\ref{p12}) degrees of initial beam polarizations in the Born and EW 1-loop approximations
  vs. cosine of the final $\tau$ lepton scattering angle at the $Z$ peak.  }
\end{figure}

In Figure~\ref{ptau_B_w_QED_EW} (left) we show the distribution of $\ptau$
in the cosine of the scattering angle at the $Z$ peak 
in the Born and 1-loop (weak, QED, and EW) approximations.
The same conventions as in previous sections are applied for the shifts $\Delta\ptau$.
The shift due to pure QED RCs is approximately a constant close to zero. 
But one can see that this observable is very sensitive to the presence 
of weak-interaction~corrections.

In the presence of initial beams polarization the expression depends on $P_{\mathrm{eff}}$:
\bqa
%\ptau(\cos\vartheta_{\tau}) \approx \frac{\ael \ata P_{\mathrm{eff}} (1+\cos^2{\vartheta_{\tau}}) 
%              -\ata (1+\cos^2{\vartheta_{\tau}}) - 2 \cos{\vartheta_{\tau}}(\ael - P_{\mathrm{eff}})}
%     {(1+\cos^2{\vartheta_{\tau}}) - \ael P_{\mathrm{eff}} (1+\cos^2{\vartheta_{\tau}}) + 2 \ata \cos{\vartheta_{\tau}} (\ael - P_{\mathrm{eff}})}.
%     \\
\ptau(\cos\vartheta) \approx -\ \frac{\ds \ata (1-\ael P_{\mathrm{eff}}) + \frac{2 \cos{\vartheta_{\tau}}}{(1+\cos^2{\vartheta_{\tau}})}(\ael - P_{\mathrm{eff}})}
     {\ds (1 - \ael P_{\mathrm{eff}}) + \frac{2\cos{\vartheta_{\tau}}    }{(1+\cos^2{\vartheta_{\tau}})}\ata(\ael - P_{\mathrm{eff}})}.     
\eqa
which can be reduced to the short form neglecting the $\ael\ata$ and $\ael P_{\mathrm{eff}}$ terms:
\bqa
\ptau(\cos\vartheta_{\tau}) \approx -\ata - \frac{2 \cos{\vartheta_{\tau}}}
     {(1+\cos^2{\vartheta_{\tau}}) }(\ael - P_{\mathrm{eff}}).
\eqa
% Sometimes in the literature it is reduced to the short approximate form by neglecting the 
%$\ael\ata$ and $\ael P_{\mathrm{eff}}$ terms:
%\bqa
%\label{ptauZpeak_ep}
%\ptau = -\ata - \frac{2 \cos{\vartheta_{\tau}}(\ael - P_{\mathrm{eff}})}
%     {(1+\cos^2{\vartheta_{\tau}}) }.
%\eqa

 The influence of the initial particle polarization on $\ptau$ at the $Z$ peak 
is demonstrated in the Figure~\ref{ptau_B_w_QED_EW} (right).
For comparison the unpolarized and two polarized cases~(\ref{p12}) as functions of  $\cos{\vartheta_\tau}$ are shown. It is seen that the behavior of $\ptau$ depends on
the polarization set choices very much, note that it even changes the sign for the $P_2$ case.
The corresponding shifts $\Delta\ptau$ also strongly depend on the initial beam polarization 
degrees and change the shape accordingly (note the maximum for $P_1$). 

\begin{figure}[h]
\centering
  \includegraphics[width=0.45\linewidth]{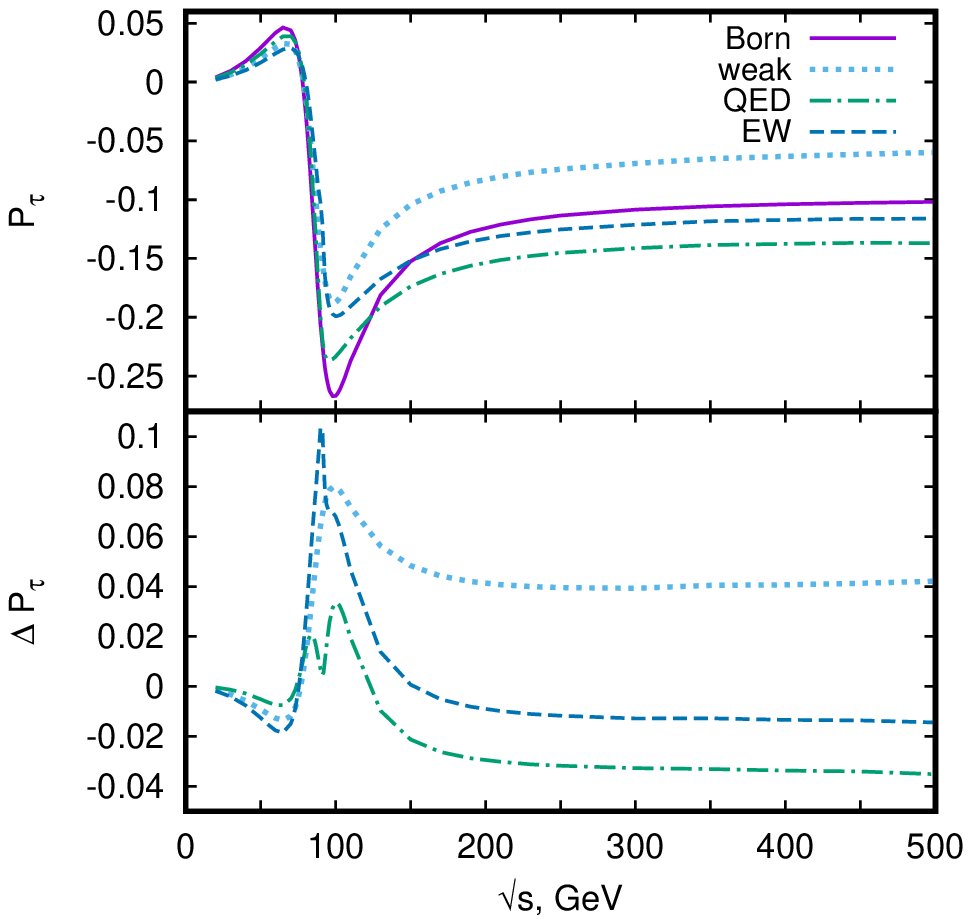}
  \includegraphics[width=0.45\linewidth]{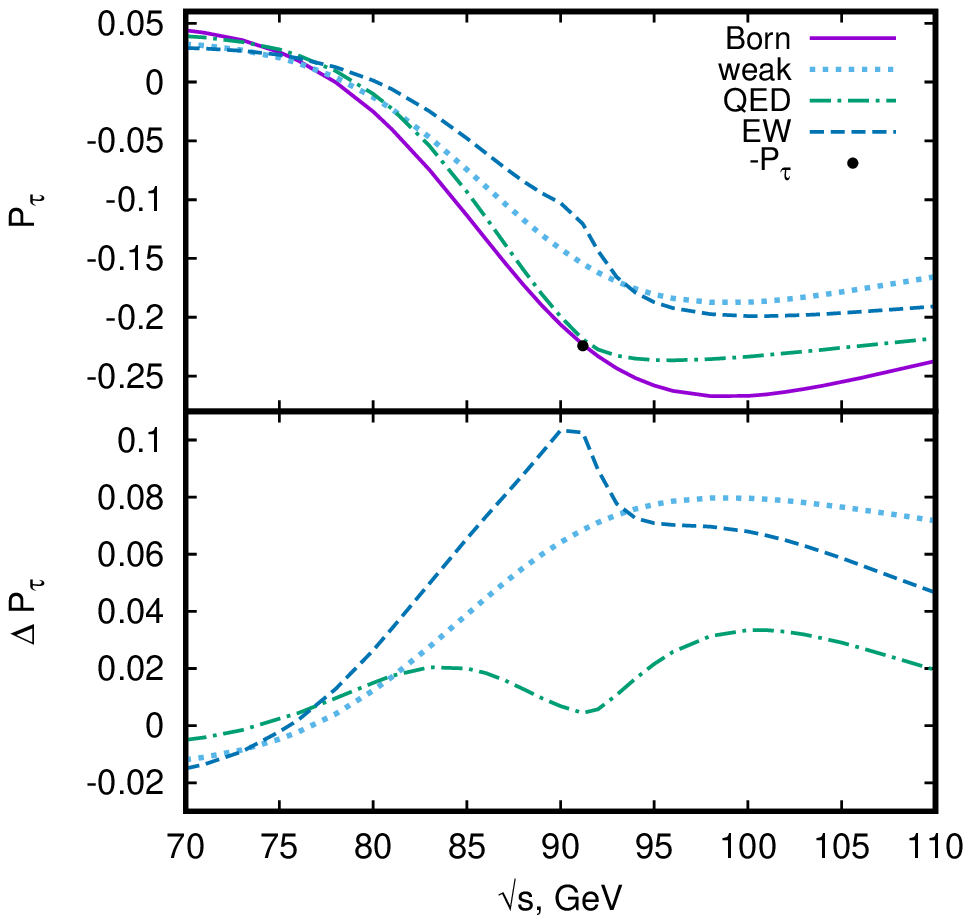}
  \caption{\label{ptau_s}
  (\textbf{Left}) The $\ptau$ polarization in the Born and 1-loop (weak, pure QED, and EW) approximations
  and $\Delta\ptau$ vs. c.m.s. energy in a wide range;
  (\textbf{Right}) the same for the $Z$ peak region.
  The black dot indicates the value $\ptau$ at the $Z$ resonance.
  }
\end{figure}

In Figure~\ref{ptau_s} we show the dependence of $\ptau$ on the c.m.s. energy
in the Born and 1-loop approximations (weak, QED, and EW).
We see that at energies above the $Z$ resonance, both weak and QED radiative 
corrections to $\ptau$ are large and considerable cancellations happen between 
their contributions. Note that theoretical uncertainties in weak and QED RCs
are not correlated, so it is necessary to take into account higher-order effects to reduce the resulting uncertainty in the complete 1-loop
result for $\ptau$ at high energies.
 
\unskip

\begin{figure}[h]
\centering
\includegraphics[width=0.55\linewidth]{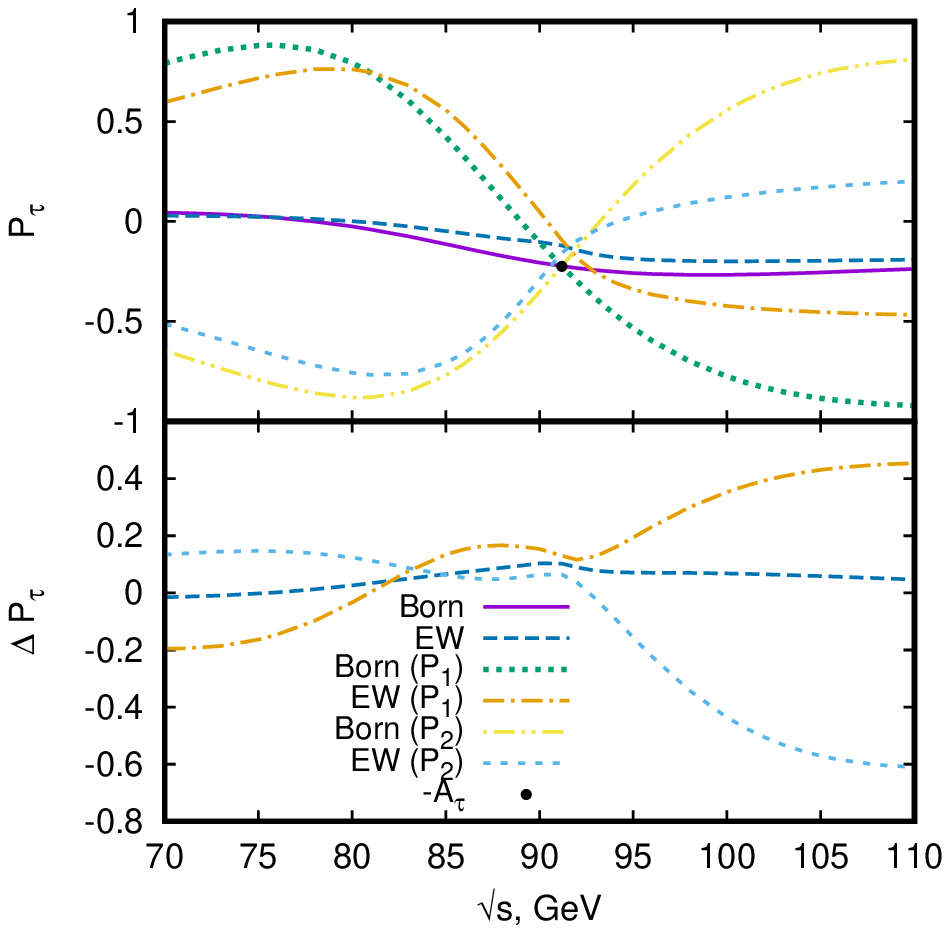}
\caption{\label{ptau_pol}
  The $\ptau$ polarization for (\ref{p12}) degrees of the initial beam polarizations 
  in the Born and 1-loop EW approximations
  vs. c.m.s. energy in the $Z$ peak region.
}
\end{figure}

In Figure~\ref{ptau_pol} we show $\ptau$ in the Born and 1-loop EW approximations 
for different sets of beam polarization degrees in a narrow bin around the $Z$ resonance.
The beam polarizations sets $P_1$ and $P_2$ are defined in Equation~(\ref{p12}).
One can see that the energy dependence of $\ptau$ is strongly affected by a beam polarization
choice outside the $Z$ peak region. The same concerns the size of radiative
corrections to $\ptau$, which are represented on the lower plot.

{\bf Summary for $\ptau$}

The $\ptau$ asymmetry is very sensitive to weak-interaction corrections 
and to the polarization degrees of the initial beams. 
Near the $Z$ resonance the value of theoretical uncertainty of $\ptau$ 
is determined by the interplay of uncertainties of
rather large contributions pure QED and weak radiative corrections.

\section{Conclusions} \label{Sect:Concl}

New opportunities of the future $e^+e^-$ colliders: GigaZ options 
and new energy scale up to several TeV require modern tools for 
high-precision theoretical calculations of observables.
We investigated  $\alr$, $\afb$ and $\alrfb$ for $e^+e^-\to \mu^+ \mu^-$ channel
and polarization $\ptau$ for the final state in $e^+e^-\to \tau^+ \tau^-$ channel
on the $Z$ resonance and in the high energy region up to 500 GeV by using {\tt MCSANC}.
We evaluated the resulting shifts of asymmetries at the Born and EW levels of accuracy
in different EW schemes.
The numerical results presented above for pure QED, weak, and complete EW radiative 
corrections show an interplay between the weak and QED contributions to asymmetries.
This fact indicates the necessity to consider those contributions always in
combined way.

Asymmetries in $e^+e^-$ annihilation processes provide a powerful tool for 
investigation of symmetries between three fermion generations.
By studying all available asymmetries, one can extract parameters 
of weak interactions in the neutral current for all three charged leptons. 
So, by comparing the parameters it will be possible to verify the 
lepton universality hypothesis at a new level of precision.

Hypothetical extra neutral $Z^\prime$ vector bosons~\cite{Langacker:2008yv} can contribute 
to the processes of $e^+e^-$ annihilation. For example, effects of Kaluza--Klein excited 
vector bosons in the gauge Higgs unification on $e^+e^-$ annihilation cross sections were 
considered in~\cite{Funatsu:2017nfm,Funatsu:2019ujy}. Since the new bosons can have 
couplings to left and right fermions being different from the SM ones, the asymmetries 
(especially with polarized beams) can help a lot in search for such $Z^\prime$ bosons.

At the FCC-ee we have experimental precision tag in the $\sin^2\vartheta^{\mathrm{eff}}_W$ 
measurement of the order of $5\times 10^{-6}$, which means more than a thirty-fold improvement %We changed the centered dot to multiplication sign
with respect to the current precision of $1.6\times 10^{-4}$. This is due to
a factor of several hundred improvement on statistical errors
and because of a considerable improvement in particle identification and vertexing.
In order to provide theoretical predictions for the considered asymmetries
with sufficiently small uncertainties
which would not spoil the precision of the future experiments besides the complete
1-loop EW radiative corrections presented here we need:
\begin{itemize}
\item higher order pure QED corrections preferably with resummation;
\item higher order (electro)weak corrections; 
\item taking into account perturbative and nonperturbative quantum chromodynamics (QCD) effects in RCs; 
\item Monte Carlo event generators and integrators which ensure the 
required technical precision.
\end{itemize}

Challenges in calculations of higher order QED effects for FCC-ee
were discussed in Ref.~\cite{Jadach:2019bye}. In Ref.~\cite{Jadach:2018lwm} it was shown 
that the uncertainty of charge asymmetry near the $Z$ peak due to the initial and final 
interference QED corrections can be reduced down to $\delta \afb \simeq 3\cdot 10^{-5}$
using exponentiation of higher-order QED effects.
The complete two-loop electroweak 
corrections in the vicinity of the $Z$ boson peak have been presented in~\cite{Dubovyk:2019szj}.
Top quark polarization in production of  $t \bar t$ pairs at a polarized linear $e^+ e^-$ collider 
was surveyed in Ref.~\cite{Groote:2010zf}. The NLO QCD corrections to the polarized decay of $Z$ 
and $W$ bosons into heavy quarks were evaluated in Ref.~\cite{Groote:2012xr} and applied to 
polarization effects in processes of electron-positron annihilation.
More details on challenges for high-precision theoretical calculations
for future $e^+e^-$ colliders can be found in~\cite{Blondel:2018mad,Blondel:2019qlh,Blondel:2019vdq}.

%%%%%%%%%%%%%%%%%%%%%%%%%%%%%%%%%%%%%%%%%%
\subsection*{Acknowledgments}
This research was funded by RFBR grant 20-02-00441.
The authors are grateful to Ya.~Dydyshka,  R.~Sadykov, 
V.~Yermolchyk, and A.~Sapronov for fruitful discussions and numerical cross checks,
and to A.~Kalinovskaya for the help with preparation of the~manuscript.

\bibliography{asymmetry_SANC_arxiv}

%%%%%%%%%%%%%%%%%%%%%%%%%%%%%%%%%%%%%%%%%%
\end{document}